# Turning genome-wide association study findings into opportunities for drug repositioning

Alexandria LAU[1] and Hon-Cheong SO[1-5]

[1] School of Biomedical Sciences, Faculty of Medicine, The Chinese University of Hong Kong, Hong Kong SAR, China

[2] KIZ-CUHK Joint Laboratory of Bioresources and Molecular Research of Common Diseases, Kunming Zoology Institute of Zoology and The Chinese University of Hong Kong, Hong Kong SAR, China

[3] Department of Psychiatry, The Chinese University of Hong Kong, Hong Kong SAR, China

[4] Margaret K.L. Cheung Research Centre for Management of Parkinsonism, The Chinese University of Hong Kong, Hong Kong SAR, China

[5] Shenzhen Research Institute, The Chinese University of Hong Kong, Shenzhen, China

*Corresponding author. Email: hcso@cuhk.edu.hk

# Abstract

Drug development is a very costly and lengthy process, while repositioned or repurposed drugs could be brought into clinical practice within a shorter time-frame and at a much reduced cost. The past decade has observed a massive growth in the amount of data from genome-wide association studies (GWAS). The rich information contained in GWAS data has great potential to guide drug discovery or repositioning. Here we provide an overview of different computational approaches which employ GWAS data to guide drug repositioning. These methods include selection of top candidate genes from GWAS as drug targets, deducing drug candidates based on drug-drug and disease-disease similarity, searching for reversed expression profiles between drugs and diseases, pathway-based methods as well as repositioning based on analysis of biological networks. Each method is illustrated with examples, and their respective strengths and limitations are discussed. Finally we discussed several areas for future research.

# Introduction

Drug development is a very costly and lengthy process, which typically involves multiple processes from drug target discovery, clinical trials to final approval by the FDA or other government agencies. The estimated cost of developing a new drug is around USD2.6 billion[1]. As a result, there have been increased interests to repurpose or reposition existing drugs for new usage. Repositioned drugs can be brought to clinical practice in a much shorter time-frame and at a much lower cost, as these drugs have gone through pharmacokinetic/pharmacodynamic and safety profiling during development. Besides existing drugs with known indications, drugs that are shelved due to failure in clinical trials may also serve as candidates for repositioning. In fact, repurposing these drugs may serve to recover the high cost that went into developing them.

In practice, many drugs in wide use today stem from repositioning. Two classical examples are Sildenafil and thalidomide[2], which are now commonly used to treat erectile dysfunction and multiple myeloma, although they were not originally designed for these indications. However, these and many other drugs are discovered based on serendipity alone while computational repositioning approaches offers a more systematic way (as well as lower cost when compared to experimental methods) of discovering such unexpected relationships between drugs and diseases.

The past decade has witnessed a massive rise in the amount of 'omics' and other forms of biomedical data (e.g. electronic health records), which makes computational approaches an attractive option to prioritize repositioning candidates. One of the fastest growing types of data comes from genome-wide association studies (GWAS), a high-throughput technique that interrogates the whole genome for common genetic variations that contribute to diseases or traits. GWAS have been highly successful in unraveling the genetic basis of many complex traits or diseases[3], and many statistical/computational methodologies have been developed to improve the power in detecting susceptibility variants. However, from a clinical point of view, one of the most important questions would be: could GWAS findings be translated into opportunities for drug discoveries or repositioning? This question also calls for more innovative approaches to analyzing GWAS data with a translational focus. In this article, we shall review several categories of methods for prioritizing drug repositioning candidates, highlighting their applications as well as their respective strengths and limitations.

# Overview of GWAS and its potential in guiding drug discoveries/ repositioning

Genome wide association studies (GWAS) aims to decipher associations between common genetic variants and traits or diseases. Typically the genetic variant studied is a single nucleotide polymorphism (SNP), and current GWAS arrays allow millions of SNPs across the whole genome to be interrogated at the same time. Variants that are not genotyped can also be imputed with appropriate reference panels[4]. The most common design for GWAS is a population-based case-control study, in which we recruit subjects with and without the disease, and search for SNPs with significant differences in allele frequencies between the two groups. However, GWAS can also be used to study continuous or time-to-event outcomes. A full GWAS workflow tutorial can be found in [5].

A number of GWAS resources are available online, and one of the largest is the GWAS catalog (https://www.ebi.ac.uk/gwas/). It is a structured repository of the summary statistics for a large variety of traits. As of Oct 14, 2019, the GWAS Catalog contains 7796 publications and 159202 associations, showing the popularity of GWAS and the vast amount of data available for mining. Other useful resources include the LD-hub (http://ldsc.broadinstitute.org/ldhub/), GWAS summary statistics from the UK Biobank (http://www.nealelab.is/uk-biobank) and dbGaP (https://www.ncbi.nlm.nih.gov/gap/) which allows access to raw genomic data to authorized users.

In order to treat a disease, it is critical to understand the causal factors, which is one of the main goals of GWAS - to understand the underlying biological mechanisms. Unlike Mendelian diseases where typically only one or few genes cause the condition, complex disease arises as a combination of a multitude of genetic and environmental causes [6]. GWAS has the niche of identifying risk loci for a disease without a priori hypotheses [7]; hence it is useful in identifying *novel* genetic loci that are beyond our current understanding of the disease. At the same time, the development of drugs with new mechanisms of action has become increasingly difficult; GWAS data therefore hold great potential in guiding drug discovery. Another important advantage is that for many diseases (e.g. psychiatric disorders[8] and cancers[9] ), current cell-based or animal models are unable to fully mimic the human condition, limiting the success rate of translating preclinical findings into clinical practice. On the other hand, GWAS are based on clinical samples of patients with actual phenotype data, and may more realistically reflect the genetic basis of the condition under study.

In an important study, Nelson et al.[10] showed that the proportion of drugs with direct genetic support increases along the development pipeline, rising from 2.0% at the preclinical stage to 8.2% among the approved drugs. In a more updated analysis by King et al.[11], they reported similar findings that genetically supported targets were more likely to be successful in Phases II and III, especially when the genes implicated are likely to be causal.

However, GWAS has been criticized for the interpretability of its findings and small effect sizes of most susceptibility variants. Approximately 90% of the SNPs found by GWAS are located in non-coding regions, suggesting that these variants may be associated with diseases through other means such as splicing, small ncRNA, lnRNA, promoter or transcription factor (TF) [6,12,13]. It is also possible that some SNPs are in linkage disequilibrium (LD) with the causal variants but are not causal themselves. However, an increasing number of methods and tools are available to uncover the functional role of SNPs and prioritize the causal genes involved, which will be discussed in detail in the sections below. Another concern is that most susceptibility variants found so far only confer small effects to diseases or traits. However, small effect sizes of individual SNPs do not necessarily dictate low efficacy when we target the corresponding protein by a drug. For instance, one of the most successful lipid lowering drugs is statins which targets HMG Co-A reductase (HMGCR). This target is also suggested by genetic evidence, since GWAS of LDL-cholesterol (LDL-C) also implicated SNPs in this loci. However, the effect size of SNP(s) at *HMGCR* is relatively modest [14,15] (e.g. G allele of rs17238484 allele was associated with ~0.06 mmol/L reduction in LDL-C), but the effect size of statins is much larger[16]. Therefore, modest effect sizes of genetic variations do not exclude therapeutic potential of the corresponding targets.

# Approaches to computational drug repositioning

## Selecting top candidate genes from GWAS as targets for drug repositioning

Perhaps the most intuitive approach to drug repositioning with GWAS is to focus on the top candidate genes identified from the study. We may first map the SNP onto corresponding genes, preferably with knowledge of the functional role of the SNP (e.g. whether it affects expression or regulation of a gene). In the next step, we may query these target genes in drug databases where information about drug-gene and drug-disease indication can be retrieved. Finally, drug repositioning opportunities present itself as a 'mismatch' between the drug indication and the disease of interest. For example, we may find a drug that targets the GWAS top gene but it has not been used for treating the disease of interest yet. The target genes retrieved from GWAS serve as a connection between the disease and drugs.

Mapping susceptibility SNPs from GWAS to the corresponding functionally important gene is a fundamental step for this approach of repositioning, and is a topic of active research [6,17–19]. However, this can be a challenging process since many SNPs are located in non-coding regions where the functional roles of variants are not fully understood. There are several comprehensive reviews on computational approaches and tools for finding relevant genes from GWAS hits [6,17–20]. We shall highlight a few approaches below and discuss the limitations of this drug repositioning methodology.

### An overview of SNP to gene mapping

Following Edwards et al [17], mapping GWAS variants to the 'target' (i.e. functionally relevant) genes can be broken down into several steps. The first step involves fine mapping of the SNPs associated with the trait of interest. Briefly, fine mapping is a process of identifying the most probable causal candidate SNP within the identified genetic loci[6,21,22] (since the GWAS chips are not designed to necessarily sequence the functional SNP, rather SNPs that are the most representative [ie 'tag' SNPs] are chosen [17]). The second step involves *in silico* methods for annotation and characterization of the functional impact of identified SNPs. Functional significance of the fine-mapped loci can be investigated with information such as chromatin accessibility, TF binding, DNA protein interactions, histone modification, DNA methylation and chromatin interactions etc. Another common approach is to look for overlap of the identified variants with expression- or other types of quantitative trait loci (QTL). Finally, if resources allow, one may perform further experiments in cell lines or animal models to ascertain the role of the target genes.

Using functional annotation to map associated SNPs to genes

Given that most associated SNPs are located in non-coding regions, it is logical to hypothesize that these SNPs may regulate gene expressions in certain ways to affect disease risks. The Encyclopedia of DNA elements (ENCODE) is one of the earliest initiatives to systematically characterize functional elements in the human genome. ENCODE aims to extensively characterize multiple genetic elements, examples of which include TF binding regions, chromatin and DNA accessibility, histone modification, epigenetics and 3D chromatin interactions. The project employs a variety of techniques including RNA-seq, DNase-seq, FAIRE-seq and ChIP-seq etc., which provide very rich data for functional annotation. Other useful resources for functional annotation include modENCODE, the NIH Roadmap Epigenomics Project, GWAS3D [23] and its successor GWAS4D [24]. One point to note is that the genes closest to the associated SNP may not necessarily be the most functionally relevant gene; a recent study suggested that the likely causative genes are often >2Mbp from the index SNP [25]. To improve the reliability of gene mapping, it is advisable to employ a variety of annotation methods to prioritize the best genes as drug target candidates.

Using expression-QTL (eQTL) to map associated SNPs to genes

Besides prioritizing the corresponding 'target' gene for the associated SNPs, it is preferable to also determine the directionality of such relationships to facilitate drug repositioning. One important question is whether the SNP causes changes in gene expression, and if so, what is the direction of change and which tissues are involved. For example, if the identified SNP causes an upregulation of gene *X* leading to increased risk of a disease, then an inhibitor of its protein product may be considered a repositioning candidate. One of the largest eQTL resources is the Genotype-Tissue Expression (GTEx) project (https://gtexportal.org/home/), which includes eQTL data from 49 tissues of over 800 subjects. However, most of the subjects are Europeans with male predominance (~67%), and the sample size may still be insufficient to detect eQTL with modest effects. Another related approach is to use tools such as PrediXcan[26] to *impute* the expression changes based on raw genotype or GWAS summary data.

## Querying drug databases for repositioning candidates

After identifying the most relevant genes from associated SNPs, finding candidate drugs which target the selected genes is relatively straightforward. For instance, in a recent attempt to repurpose drugs for inflammatory bowel disease (IBD), Grenier et al [27] first selected the most likely causal SNPs by Bayes factor within each locus, then these SNPs were mapped to relevant genes by functional annotation. Finally, drugs that may be repositioned for treating IBD (if not already known) were derived from these genes using a web-based tool Gene2Drug [28] (see the section: Using pathway analysis in combination with other approaches). Other reports pulled drug data from multiple databases including TTD, Pandrugs, PharmGKB and DrugBank to derive drug-target relationships [29–35].

## Limitations

There are several limitations of using the top candidates genes for direct drug repositioning. Firstly, the top genes identified from GWAS may not be easily druggable. Finan et al. performed a comprehensive analysis on the druggability of genes [36]. They estimated that only 4,479 (22%) of the 20,300 protein coding genes are druggable (or already targeted by a drug). Second, focusing on the effect of the top SNPs may miss biologically meaningful target genes with small effect sizes [37–39]. Third, focusing on single candidate gene may miss multi-target drugs, which could be more effective than single-target ones for some conditions [40]. Also, recently increasing attention has been placed on development of multi-target drugs [41,42]. Fourth, as discussed earlier, due to the complexity of the human genome, there is no perfect way to proper annotation. As a result, different studies may have employed different (sometimes incomplete) annotation procedures, and integrating various annotation approaches is not straightforward.

Table 1: Summary and comparison of drug repositioning approaches using GWAS data

| Approach | Brief description | Pros | Cons | Selected references |
|---|---|---|---|---|
| Using functional annotation for identifying top | Map GWAS SNPs to corresponding functionally related genes with | Relatively clear biological interpretation; straightforward | Usually only single or a few genes are examined; may potentially miss | [6,17,19] |

| candidate genes (then link to drugs) | functional annotation tools | computation and low computational cost; multiple databases available for annotation | multi-target drugs; directionality of effect may not be clear; functional annotation information for some SNPs may be missing; not all genes are directly druggable | |
|---|---|---|---|---|
| Using eQTL data for identifying top candidate genes (then link to drugs) | Map GWAS SNPs to corresponding gene using eQTL information | Directionality indicated; covers SNPs that affects multiple genes at the same time; easy to implement | GTEx data still limited in sample size, and bias towards European population; not all SNPs may affect gene expression | [43–45] |
| Pathway/gene-set analysis | Repositioning drug based on pathway or gene-set analysis of GWAS results | Consideration of drug effect on a genome-wide scale; multi-target drugs included; inclusion of risk loci with small effect size individually but good therapeutic potential when combined as a pathway | Definition of pathways can be complicated; incomplete characterization of pathways for all drugs; directionality of effect may not be clear | [38,46] |
| Similarity-based: Drug-drug or disease-disease similarity | Evaluating similarity between drugs and diseases effect may reveal novel drug-disease relationships | Intuitive in concept; simple computation; less detailed understanding of drug mechanism required | 'Similarity' may not be easy to define; difficulty in integrating different sources of similarity measures; relatively hard to uncover drugs with novel mechanisms of actions | [47,48] |
| Reversed expression pattern between drugs and diseases | A drug with expression profile opposite to that of a disease are candidate therapeutic agents | Considers data across many genes instead of the most significant ones; imputed expression readily available for many tissues and from large GWAS samples, and less susceptible to confounding and reverse causality; understanding of drug mechanism not required; relatively better at uncovering drugs of novel mechanism | Expression reversal may not be the only drug mechanism; limitation of cell lines (cannot fully model human conditions); imputation accuracy of some genes may be poor | [49,50] |

| Network-based methods | Integration of multiple sources of data regarding drugs, proteins, genes and diseases relationships to reveal novel drug-disease connections | Flexible; ability to integrate multiple sources of data; well established network analysis methods from other fields | Integrating data with different nature and potentially different kinds of bias is difficult; complicated parameter optimization; difficulty in determining edge strength; relatively less capable of revealing unexpected repositioning candidates | [51–55] |
|---|---|---|---|---|

## Drug repositioning based on pathway or gene-set analysis

Pathway or gene-set analysis (as opposed to single gene or SNP-based studies) offers a more macroscopic view of the biological processes underlying diseases and drug effects. The key idea behind pathway analysis is to organize various functionally or biologically relevant genes together and consider the overall effect. As mentioned earlier, approaches that consider only individual SNPs or genes may miss biologically meaningful associations of modest effect sizes [56]. Biological pathways group genes based on their function; on the other hand, 'gene-sets' can be any set of functionally related genes or a set based on arbitrary criteria set by the researcher (in this context, drug targets or 'effector genes'). For the purpose of this review, we refer to both of these grouping methods as 'pathways'. Several examples of applications are discussed below.

De Jong et al. [48] studied repositioning candidates for schizophrenia by a gene-set analysis based on GWAS summary statistics, where each drug pathway is defined by pharmacological profiles and chemical binding affinity. The analysis highlighted several candidates reaching a suggestive level of significance, including two dopamine receptor antagonists and a tyrosine kinase inhibitor. In another recent work, So et al. [57] investigated whether findings from GWAS may be used to guide drug repositioning for depression and anxiety disorders. Drug-effector gene-sets were extracted with DSigDB and gene-based significance from GWAS was computed by FASTBAT [58]. Then pathway analyses (following the principle of MAGMA[59]) were conducted to look for enrichment of GWAS results for specific drugs. Interestingly, the repositioning hits identified are largely enriched for known psychiatric drugs or those included in clinical trials. Enrichment was seen for antidepressants and anxiolytics but also for antipsychotics. The study also revealed other repositioning candidates with literature support. Note that the above two studies did not only focus on the top genes but also considered the actual significance level of each gene.

Another approach was presented by Jhamb et al. [60]. The authors first obtained GWAS data from STOPGAP [61]. After SNPs with *p*<5e-8 are mapped to genes, these genes were mapped to their respective pathways with MetaBase, a manually curated software suite that contains interaction data. The main novelty is that the authors considered an expanded set of GWAS 'hits' by including genes in related pathways as well. Finally a 2 by 2 table is constructed for each disease to look for over-representation of the (expanded) GWAS hits among drug targets indicated for each disease.

Yet another work by Gaspar et al. investigated drug repositioning opportunities for schizophrenia using GWAS data [62]. They also proposed a new visualization approach and studied how increased sample size of the original GWAS may improve the yield of drug repositioning. Pathway analysis can also be combined with other methods. Mäkinen et al. [37] first mapped cardiovascular disease (CVD) associated-SNPs to genes with eQTL, then investigated the enrichment of these 'e-SNPs' among known biological pathways. After obtaining the CVD pathway set, the pathways are augmented with co-expression analysis and subsequently placed in a Bayesian network model of gene-gene interactions. By integrating a larger variety of sources of omics data, the identified genes may serve as better targets for repositioning.

Another tool combining pathway analysis and perturbation library is Gene2Drug [28], which is used by Grenier et al. [27] (see discussion above). Gene2Drug takes a single gene as input, and aims to discover drugs linked to the input gene. The program extracts drug-related pathways based on perturbation data from CMap (see section below), then generates the subset of pathways including the input gene.

Pathway-based analysis is a flexible approach and several extensions are possible. For example, in clinical practice comorbid diseases are very common, and it will be preferable to find drugs that target both diseases at the same time. A recent study proposed a pathway-based approach to address this problem. Wong et al.[63] first employed a false discovery rate (FDR)-based approach to uncover genetic loci *shared* between depression/anxiety and cardiometabolic diseases from GWAS summary data. The shared loci were then subject to pathway analysis to uncover repositioning candidates, many of which are supported by the literature. Also, while the focus of this review is on repositioning using GWAS data, the pathway-based approach can be readily extended to handle other types of human genomic data. For example, a recent study showed by pathway analysis that *de novo* mutations may be used to guide drug discoveries for neuro-psychiatric disorders[64]. Readers may also refer to [65] for further discussions on the potential of GWAS data in drug discovery for neuro-psychiatric disorders.

### Limitations

First, defining a pathway is complicated because feedback inhibition and compensatory mechanism is almost ubiquitous in biology [38]. Also, there is an emphasis on the known pathways in pathway-based approaches [60], therefore the applicability of this approach may be limited by our current knowledge of biology. The mechanisms of many drugs are not entirely clear so their 'pathways' or gene-sets may be incompletely characterized. Finally, the directionality of effect may not be clear.

## **Similarity-based drug repositioning**

Similarity between drug-drug and disease-disease pairs may be leveraged for drug repositioning. One of the characteristics of such similarity-based approaches is that they generally do not require detailed understanding of drug or disease mechanisms, and are able to consider high-throughput omics data across many genes without restricting to the most significant ones. Note that similarity-based approaches is closely related to network-based approaches, however some algorithms have been specifically designed for the former and network-based methods usually involve modeling a larger variety of information (apart from similarity between drugs and diseases).

### Drug-drug similarity match

Drug-drug similarity analysis begins with the comparison between the chemical and/or biological profiles of different drugs, after which drugs can be repositioned to the indications of another one if the two shared sufficient similarity. An example to illustrate this idea is the work by Napolitano et al.[66]. The authors first generated a drug similarity matrix by integrating multiple sources of information, such as similarity in transcriptomic changes after drug administration (from CMap), as well as similarity in drug target proteins and chemical structures. The authors integrated different sources of information via a machine learning (ML) framework and proposed candidates for repositioning based on similarity to the known drugs.

### Disease-disease similarity match

It is reasonable to hypothesize that if two diseases are similar enough, then the drugs that are indicated for one disease may be repositioned to treat the other. This concept has been used in a number of studies in drug repositioning. For instance, Gottlieb et al. [67] proposed an algorithm (PREDICT) to combine multiple measures of drug-drug similarity and disease-disease similarity for repositioning. Later Wang et al. [68] combined the approach in PREDICT with drug and disease data from DrugBank, CMap, OFFSIDE (a side effect database from PharmGKB) and literature text mining to create a recommendation system for novel drug candidates. While the aforementioned studies did not directly utilize GWAS data, such data may be used to measure disease-disease similarity as well. For example, as a proof of concept, Li et al. [47] examined pairs of genetic variants shared between disease while considering eQTL information at the same time. They found that pairs of diseases that are genetically similar were also more likely to be comorbid in clinical samples. The study suggests genetic data from GWAS may also be used to assess 'similarity', which might have implications in drug repositioning.

Limitations

Similarity between drugs and diseases may not be easy to define. For example, some studies used CMap to ascertain drug-drug similarity, but there are limitations of the CMap database (please also refer to discussions in the next section), such as lack of appropriate tissue-specific cell lines in some occasions. How to integrate different sources of data to define similarity remains an open question. By its nature, this approach may tend to uncover candidates that are similar to the known drugs, and is less capable of revealing drugs with novel mechanisms of actions. It is possible that the drug candidates is already suggested by clinical experience due to their similarity with the known ones. Network-based approaches may also share similar shortcomings, due to similarity in their principles.

# **Searching for reversed expression profiles between drugs and diseases**

A more distinct approach to repositioning is to compare the expression profiles of drugs against those of specific diseases. The core hypothesis is that if a drug produces an expression profile that is opposite to that of a disease, then the drug may be considered a repositioning candidate (due to its potential to 'reverse' disease-related expression profiles). This approach does not require knowledge of the drug mechanisms or even drug targets; it can be applied as long as drug-induced expression profiles are available. In addition, it utilizes GWAS data across multiple genes instead of focusing on only the top significant ones. As will be described later, the method is also readily available to any complex diseases/traits with GWAS summary statistics available. Compared to similarity or network-based methods, this approach may have a greater chance of uncovering drugs with new mechanisms of action.

To perform this kind of analysis, one would need the following components: (1) Drug-induced (differential) expression profile; (2) Disease-induced (differential) expression profile; and (3) algorithms to determine the correlations between (1) and (2).

Publicly available data are available for drug-induced expression profiles. Popular choices include the Connectivity Map (CMap) [69] and its successor Library of Integrated Network-based Cellular Signatures (LINCS) L1000 data [70]. CMap is a collection of expression profiles by applying ~1300 compounds on 5 human cancer cell lines, with expression profiles for ~7000 genes as of Build 2. LINCS (L1000) is a database of much larger scale that contains about 1328098 gene expression profiles as a result from the applications of ~42553 perturbagens. A set of 1000 'landmark' transcripts was directly measured while the expression levels of other genes were mainly computed by imputation. A comprehensive review about these resources is given by Musa et al. [71].

For the purpose of this review, we shall mainly discuss GWAS-derived data for component (2). While one may use RNA-seq or microarray data in patients to directly estimate component (2), as argued in [50], there are several advantages of using imputed expression profile from GWAS. For example, patients are often medicated before their samples are collected, which may affect their expression profiles. GWAS-imputed transcriptome is much less susceptible to confounding by medication or other environmental factors. In addition, expression can be imputed for a large variety of tissues, even for those (e.g. brain) which are difficult to access. The sample size of GWAS is also often much larger than standard RNA-seq or microarray studies on clinical samples. This approach of comparing GWAS-imputed expression against drug transcriptomes was proposed and adopted by So et al. [50] to prioritize repositioning candidates for psychiatric disorders. The method revealed numerous candidates supported by previous preclinical or clinical studies, and the approach was able to 're-discover' known psychiatric medications for the respective disorders, despite having no prior knowledge of the drug indications.

Imputation of gene expression profiles could be performed by various tools, the most popular being PrediXcan [26] or its successor S-PrediXcan [72]. The latter is able to impute expression changes by GWAS summary statistics alone, which enhances the applicability of the above repositioning methodology as summary statistics are now widely available. Briefly, these tools first learn how SNPs may dictate gene expression levels from training data extracted from datasets such as GTEx, GEUVADIS and DGN. Based on the prediction model built, the tools can then predict or impute expression levels when presented with genotypes from new samples. Finally, for component (3) of the analysis, we need to assess the (anti-)correlation between the expression profiles from drugs and diseases. In particular, So et al. [50] adopted Spearman and Pearson correlation as well as Kolmogorov–Smirnov (KS) test to evaluate patterns of reversed expression.

Limitations

There are several limitations of this approach, many of which are also discussed in [50]. First, perturbation libraries such as CMap and L1000 are usually based on cell line data, meaning that all the limitations of cell line-based experiments may also apply. For instance, cell lines cannot model complex cell-cell or cell-ECM (extracellular matrix) interactions, and pharmacokinetic properties and dosage control are often experiment protocol-dependent [73]. Second, many of the cell lines are cancer cell lines; using them for expression profiling may not be optimal when the aim is to prioritize drugs for other diseases, such as psychiatric disorders or CVD. Third, not all therapeutically important drugs work by reversing gene expression profiles, although the approach appears to work reasonably well at least in psychiatric disorders. Finally, imputation of expression profiles from GWAS data may not be as accurate as directly measuring the actual tissue-specific expression profiles from patient samples, although the latter is often difficult to collect. Some genes may be poorly imputed due to large environmental influence on gene expression [74], or that the original sample is not powerful enough to detect the eQTLs. Limitations of the GTEx sample discussed earlier also apply here.

# Network-based approaches

## Overview

Network-based drug repositioning approaches aim to uncover novel drug-disease relationships by integrating a wide variety of biological information. A typical network consists of nodes (genes, protein, disease/trait, compounds) and edges (often weighted in biological settings) connecting the nodes. This approach is a popular and well-established drug repositioning technique which offers high flexibility, as it allows for consideration of multiple dimensions of data sources. The types of biological networks that are useful for drug repositioning include, for example, gene regulatory, gene-gene interaction, metabolic, drug-target interaction (DTI) and protein-protein interaction (PPI) [53] networks. It is preferable to integrate multiple biological networks in order to reduce noise and improve biological relevance [52,54]. After preparing a curated network of information, analyzing the graphs usually involves tools developed from graph theory. For drug repositioning, two types of network analysis approaches might be useful: clustering and propagation [75]. Briefly, clustering is a way of discovering subnetworks by the similarity of its elements, since biologically related entities should intuitively share a handful of underlying connections [76,77]. Clustering may reveal subnetworks and new relationships between drugs and diseases, fostering the discovery of drug candidates. Propagation approaches, as the name suggests, models the propagation of information from a source node to its surroundings [78]. Approaches such as random walk[79] may be used to assess the distance between a drug and a disease, thereby prioritizing drug candidates. For the purpose of this study, we shall not focus on discussing the general approaches of network-based drug repositioning which has already been extensively reviewed [51–55,80–83]. Instead, we shall focus on the role and contribution of GWAS data towards this type of drug repositioning.

## Using GWAS data in network-based drug repositioning

GWAS data can provide richer information for disease-disease, disease-gene or drug-gene edges in biological networks, hence providing better repositioning candidates or new drug targets. In a recent work, Gaspar et al. built bipartite drug-target networks leveraging gene-based statistics from MAGMA and S-PrediXcan, accounting for both SNP-level associations and imputed transcriptomic changes. They also built an online tool Drug Targetor ([drugtargetor.com](drugtargetor.com)) which visualizes the resulting drug-target network. The authors built a network from GWAS of major depressive disorder, and suggested potential new drug candidates and their modes of actions [84].

In another application, a network-based approach was used to analyze GWAS results of CVD [85]. The study began with pathway analysis of the significant SNPs from 16 GWAS datasets. The identified pathways were then mapped to the PPI network InWeb and analyzed using random walk. Next, the authors examined the topological properties of the identified clusters and prioritized CVD-associated genes that displayed high centrality and betweenness, which may be prioritized as potential drug targets. In another report, Shu et al. [86] studied shared genetic networks and key 'driver genes' for both CVD and T2D by a systems approach. They considered co-expression modules for CVD and type 2 diabetes (T2D), and incorporated gene regulatory networks from GIANT [87] and Bayesian networks constructed from CVD and T2D related tissues to prioritize key drivers. The potential driver genes or key regulators were further validated by cell culture and animal models. These genes served as useful targets for drug discovery or repositioning. While the above two examples were not solely aimed at drug repositioning, they showed how network-based methods may shed light on biological processes and help prioritize drug target genes.

## Limitations

First, integrating different sources of networks can be tedious and challenging in terms of data cleaning, and this might require certain degree of understanding of the data and their biological meanings. Second, network analysis heavily rely on the similarity or closeness between different entities [55], therefore the repositioning results may be concentrated around the 'nearby pharmacological space'; it may be relatively difficult to uncover drugs with novel mechanisms. Also, the strengths of similarity between nodes (i.e. edge strengths) are often hard to define. In addition, limitations in data sources could affect the performance of network-based analysis methods, for example the reliability and completeness of interactome data [88] and the difficulty in constructing accurate PPI networks [89]. Finally, the complicated nature of network structures makes parameter optimization a key issue when analyzing the data with network-based algorithms [90].

# Conclusion and Future Directions

Despite more than 7000 GWAS has been conducted to date, comparatively few studies have systematically analyzed the potential of GWAS in drug discovery or repositioning. With growing resources from biobanks (e.g. the UK Biobank) and increasing availability of GWAS data, such kind of data should provide a very rich resource for guiding drug discovery/repositioning. Indeed, as discussed above, a number of studies have provided early evidence that human genomics data from GWAS might improve the success of drug development, or that repositioning strategies based on GWAS are able to 're-discover' known drugs for diseases and/or suggest reasonable new candidates.

We focus on computational approaches to *prioritize* drug candidates in this review, but we should emphasize that further experimental and clinical studies are necessary to confirm the findings. Computational or bioinformatics approaches helps to narrow down the search space and improve the success rate of development by prioritizing the best candidates, but they are not designed to provide confirmatory evidence. Nevertheless, given the huge cost and long time involved in developing a new drug, even a tiny improvement in success rate would translate to very substantial savings in absolute terms.

We hereby make a few comments to highlight several general limitations and directions for further research. First, to improve the accuracy of GWAS-based drug-repositioning studies, there is a need for more high-quality data, including but not limited to GWAS with large cohort sizes and richer phenotypes, GTEx (or similar projects) on more tissues and larger sample sizes, perturbation libraries that involve more complex drug testing models etc. Second, new statistical methods have emerged to ascertain not only associations but also *causal* relationships between exposure (or risk factors) and outcomes; one of the most prominent methods is Mendelian Randomization (MR) [91]. A few studies have suggested that MR may be used to model the effects or side-effects of drugs (with known targets). This may have implications for drug discovery or repositioning although further development in methodology may be required[92–94]. Third, machine learning (ML) and artificial intelligence are among the fastest growing areas in recent years. ML approaches such as deep learning and other methods hold great promise for accelerating drug discovery/repositioning, as they may be able to discover and predict with higher accuracy the complex patterns and relationships between genes, drugs and diseases [95–98]. For example, ML methods may be used to capture complex relationships between drug transcriptome and the drug's treatment potential for specific diseases [99]. For diseases with high heterogeneity, one type of drug may only be useful for a subgroup of patients[100]. Unsupervised learning methods may help to subtype diseases more accurately, enhancing the success rate of drug development [101]. Finally, GWAS is a very rich source of omics data, but integration with other forms of human genomic data, such as those from exome sequencing, transcriptomics and epigenomics studies will further improve the reliability of repositioning. In the same vein, another important direction is the integration of large-scale electronic health records (EHR) with multi-omics data in drug development. As emphasized in this review, each repositioning method has its own strengths and limitations. An important future direction is to integrate different methodologies for optimal prediction of drug candidates.

**Acknowledgements**
H.C.S. was supported partially by the Lo Kwee Seong Biomedical Research Fund, a Direct Grant from The Chinese University of Hong Kong, RGC Collaborative Research Fund C4054-17WF, and HMRF grant (06170506).

**Conflicts of interest**
The authors declare no competing interests.

**Legends (Figure 1)**
1. Risk loci for a particular disease can be identified from GWAS.
2. Several approaches to drug repositioning using GWAS data are highlighted:
(a) Selection of top candidate genes: the risk loci from GWAS data can be mapped to the most likely relevant genes with functional annotation and eQTL data. If the identified candidate gene is druggable and the drug is not already indicated for the disease, the drug may serve as a repositioning candidate.
(b) Pathway or gene-set analysis approach: the identified candidate gene(s) are placed in the context of its pathways; drugs that target members of the same pathway are potential drug candidates. Alternatively, the entire set of GWAS data may be used to derive gene-based statistics, and enrichment test performed to look for drugs whose targets/effector genes achieve higher significance (lower p-values) than expected as a whole.
(c) Looking for reversed expression patterns between drugs and diseases: The core hypothesis is that if a drug produces an expression profile that is opposite to that of a disease, then the drug may be considered a repositioning candidate (due to its potential to 'reverse' disease-related expression profiles).
(d) Network-based analysis: Integration of multiple sources of data such as drugs, proteins, genes and diseases relationships to construct biological networks. Further analysis with computational methods such as random walk may reveal novel drug-disease connections. The principle of this method is close to 'similarity-based' methods (see main text) for drug repositioning, but network-based method usually integrates a greater variety of information.
3. Finally, drugs can be prioritized as repositioning candidates and verified in further experimental and clinical studies.

# Bibliography

[1]  DiMasi JA, Grabowski HG, Hansen RW. Innovation in the pharmaceutical industry: New estimates of R&D costs. J Health Econ 2016;47:20–33. https://doi.org/10.1016/j.jhealeco.2016.01.012.

[2]  Shim JS, Liu JO. Recent Advances in Drug Repositioning for the Discovery of New Anticancer Drugs. Int J Biol Sci 2014;10:654–63. https://doi.org/10.7150/ijbs.9224.

[3]  Visscher PM, Wray NR, Zhang Q, Sklar P, McCarthy MI, Brown MA, et al. 10 Years of GWAS Discovery: Biology, Function, and Translation. Am J Hum Genet 2017;101:5–22. https://doi.org/10.1016/j.ajhg.2017.06.005.

[4]  Das S, Abecasis GR, Browning BL. Genotype Imputation from Large Reference Panels. Annu Rev Genomics Hum Genet 2018;19:73–96. https://doi.org/10.1146/annurev-genom-083117-021602.

[5]  Marees AT, de Kluiver H, Stringer S, Vorspan F, Curis E, Marie-Claire C, et al. A tutorial on conducting genome-wide association studies: Quality control and statistical analysis. Int J Methods Psychiatr Res 2018;27. https://doi.org/10.1002/mpr.1608.

[6]  Gallagher MD, Chen-Plotkin AS. The Post-GWAS Era: From Association to Function. Am J Hum Genet 2018;102:717–30. https://doi.org/10.1016/j.ajhg.2018.04.002.

[7]  Need AC, Goldstein DB. Whole genome association studies in complex diseases: where do we stand? Dialogues Clin Neurosci 2010;12:37–46.

[8]  Nestler EJ, Hyman SE. Animal models of neuropsychiatric disorders. Nat Neurosci 2010;13:1161–9. https://doi.org/10.1038/nn.2647.

[9]  Mak IW, Evaniew N, Ghert M. Lost in translation: animal models and clinical trials in cancer treatment. Am J Transl Res 2014;6:114–8.

[10] Nelson MR, Tipney H, Painter JL, Shen J, Nicoletti P, Shen Y, et al. The support of human genetic evidence for approved drug indications. Nat Genet 2015;47:856–60. https://doi.org/10.1038/ng.3314.

[11] King EA, Davis JW, Degner JF. Are drug targets with genetic support twice as likely to be approved? Revised estimates of the impact of genetic support for drug mechanisms on the probability of drug approval.: Supplementary Methods And Results. Genetics; 2019. https://doi.org/10.1101/513945.

[12] Schaub MA, Boyle AP, Kundaje A, Batzoglou S, Snyder M. Linking disease associations with regulatory information in the human genome. Genome Res 2012;22:1748–59. https://doi.org/10.1101/gr.136127.111.

[13] Maurano MT, Humbert R, Rynes E, Thurman RE, Haugen E, Wang H, et al. Systematic Localization of Common Disease-Associated Variation in Regulatory DNA. Science 2012;337:1190–5. https://doi.org/10.1126/science.1222794.

[14] Teslovich TM, Musunuru K, Smith AV, Edmondson AC, Stylianou IM, Koseki M, et al. Biological, clinical and population relevance of 95 loci for blood lipids. Nature

[15] Swerdlow DI, Preiss D, Kuchenbaecker KB, Holmes MV, Engmann JEL, Shah T, et al. HMG-coenzyme A reductase inhibition, type 2 diabetes, and bodyweight: evidence from genetic analysis and randomised trials. The Lancet 2015;385:351–61. https://doi.org/10.1016/S0140-6736(14)61183-1.

[16] Law MR. Quantifying effect of statins on low density lipoprotein cholesterol, ischaemic heart disease, and stroke: systematic review and meta-analysis. BMJ 2003;326:1423–0. https://doi.org/10.1136/bmj.326.7404.1423.

[17] Edwards SL, Beesley J, French JD, Dunning AM. Beyond GWASs: Illuminating the Dark Road from Association to Function. Am J Hum Genet 2013;93:779–97. https://doi.org/10.1016/j.ajhg.2013.10.012.

[18] Pritchard JE, O'Mara TA, Glubb DM. Enhancing the Promise of Drug Repositioning through Genetics. Front Pharmacol 2017;8:896–896. https://doi.org/10.3389/fphar.2017.00896.

[19] Miller JE, Veturi Y, Ritchie MD. Innovative strategies for annotating the "relationSNP" between variants and molecular phenotypes. BioData Min 2019;12:10. https://doi.org/10.1186/s13040-019-0197-9.

[20] Nishizaki SS, Boyle AP. Mining the Unknown: Assigning Function to Noncoding Single Nucleotide Polymorphisms. Trends Genet 2017;33:34–45. https://doi.org/10.1016/j.tig.2016.10.008.

[21] Schaid DJ, Chen W, Larson NB. From genome-wide associations to candidate causal variants by statistical fine-mapping. Nat Rev Genet 2018;19:491–504. https://doi.org/10.1038/s41576-018-0016-z.

[22] Benner C, Havulinna AS, Järvelin M-R, Salomaa V, Ripatti S, Pirinen M. Prospects of Fine-Mapping Trait-Associated Genomic Regions by Using Summary Statistics from Genome-wide Association Studies. Am J Hum Genet 2017;101:539–51. https://doi.org/10.1016/j.ajhg.2017.08.012.

[23] Li MJ, Wang LY, Xia Z, Sham PC, Wang J. GWAS3D: detecting human regulatory variants by integrative analysis of genome-wide associations, chromosome interactions and histone modifications. Nucleic Acids Res 2013;41:W150–8. https://doi.org/10.1093/nar/gkt456.

[24] Huang D, Yi X, Zhang S, Zheng Z, Wang P, Xuan C, et al. GWAS4D: multidimensional analysis of context-specific regulatory variant for human complex diseases and traits. Nucleic Acids Res 2018;46:W114–20. https://doi.org/10.1093/nar/gky407.

[25] Brodie A, Azaria JR, Ofran Y. How far from the SNP may the causative genes be? Nucleic Acids Res 2016;44:6046–54. https://doi.org/10.1093/nar/gkw500.

[26] Gamazon ER, Wheeler HE, Shah KP, Mozaffari SV, Aquino-Michaels K, Carroll RJ, et al. A gene-based association method for mapping traits using reference transcriptome data. Nat Genet 2015;47:1091–8. https://doi.org/10.1038/ng.3367.

[27] Grenier L, Hu P. Computational drug repurposing for inflammatory bowel disease using genetic information. Comput Struct Biotechnol J 2019;17:127–35. https://doi.org/10.1016/j.csbj.2019.01.001.

[28] Napolitano F, Carrella D, Mandriani B, Pisonero-Vaquero S, Sirci F, Medina DL, et al. gene2drug: a computational tool for pathway-based rational drug repositioning. Bioinformatics 2018;34:1498–505. https://doi.org/10.1093/bioinformatics/btx800.

[29] Rastegar-Mojarad M, Ye Z, Kolesar JM, Hebbring SJ, Lin SM. Opportunities for drug

repositioning from phenome-wide association studies. Nat Biotechnol 2015;33:342–5. https://doi.org/10.1038/nbt.3183.

[30] Sanseau P, Agarwal P, Barnes MR, Pastinen T, Richards JB, Cardon LR, et al. Use of genome-wide association studies for drug repositioning. Nat Biotechnol 2012;30:317–20. https://doi.org/10.1038/nbt.2151.

[31] Wang Z-Y, Zhang H-Y. Rational drug repositioning by medical genetics. Nat Biotechnol 2013;31:1080–2. https://doi.org/10.1038/nbt.2758.

[32] Grover MP, Ballouz S, Mohanasundaram KA, George RA, H Sherman CD, Crowley TM, et al. Identification of novel therapeutics for complex diseases from genome-wide association data. BMC Med Genomics 2014;7:S8. https://doi.org/10.1186/1755-8794-7-S1-S8.

[33] Grover MP, Ballouz S, Mohanasundaram KA, George RA, Goscinski A, Crowley TM, et al. Novel therapeutics for coronary artery disease from genome-wide association study data. BMC Med Genomics 2015;8:1–11. https://doi.org/10.1186/1755-8794-8-S2-S1.

[34] Quan Y, Liu M-Y, Liu Y-M, Zhu L-D, Wu Y-S, Luo Z-H, et al. Facilitating Anti-Cancer Combinatorial Drug Discovery by Targeting Epistatic Disease Genes. Molecules 2018;23:736. https://doi.org/10.3390/molecules23040736.

[35] Piñeiro-Yáñez E, Reboiro-Jato M, Gómez-López G, Perales-Patón J, Troulé K, Rodríguez JM, et al. PanDrugs: a novel method to prioritize anticancer drug treatments according to individual genomic data. Genome Med 2018;10:41. https://doi.org/10.1186/s13073-018-0546-1.

[36] Finan C, Gaulton A, Kruger FA, Lumbers RT, Shah T, Engmann J, et al. The druggable genome and support for target identification and validation in drug development. Sci Transl Med 2017;9:eaag1166. https://doi.org/10.1126/scitranslmed.aag1166.

[37] Mäkinen V-P, Civelek M, Meng Q, Zhang B, Zhu J, Levian C, et al. Integrative Genomics Reveals Novel Molecular Pathways and Gene Networks for Coronary Artery Disease. PLOS Genet 2014;10:e1004502. https://doi.org/10.1371/journal.pgen.1004502.

[38] White MJ, Yaspan BL, Veatch OJ, Goddard P, Risse-Adams OS, Contreras MG. Strategies for Pathway Analysis Using GWAS and WGS Data. Curr Protoc Hum Genet 2019;100:e79. https://doi.org/10.1002/cphg.79.

[39] Sakagami M. Systemic delivery of biotherapeutics through the lung: opportunities and challenges for improved lung absorption. Ther Deliv 2013;4:1511–25. https://doi.org/10.4155/tde.13.119.

[40] Talevi A. Multi-target pharmacology: possibilities and limitations of the "skeleton key approach" from a medicinal chemist perspective. Front Pharmacol 2015;6. https://doi.org/10.3389/fphar.2015.00205.

[41] Bang S, Son S, Kim S, Shin H. Disease Pathway Cut for Multi-Target drugs. BMC Bioinformatics 2019;20:74. https://doi.org/10.1186/s12859-019-2638-3.

[42] Lu J-J, Pan W, Hu Y-J, Wang Y-T. Multi-target drugs: the trend of drug research and development. PloS One 2012;7:e40262. https://doi.org/10.1371/journal.pone.0040262.

[43] Huang QQ, Ritchie SC, Brozynska M, Inouye M. Power, false discovery rate and Winner's Curse in eQTL studies. Nucleic Acids Res 2018;46:e133–e133. https://doi.org/10.1093/nar/gky780.

[44] Heinig M. Using Gene Expression to Annotate Cardiovascular GWAS Loci. Front Cardiovasc Med 2018;5. https://doi.org/10.3389/fcvm.2018.00059.

[45] Zeng B, Lloyd-Jones LR, Holloway A, Marigorta UM, Metspalu A, Montgomery GW, et

al. Constraints on eQTL Fine Mapping in the Presence of Multisite Local Regulation of Gene Expression. G3 Genes Genomes Genet 2017;7:2533–44. https://doi.org/10.1534/g3.117.043752.

[46] Pan Y, Cheng T, Wang Y, Bryant SH. Pathway Analysis for Drug Repositioning Based on Public Database Mining. J Chem Inf Model 2014;54:407–18. https://doi.org/10.1021/ci4005354.

[47] Li H, Fan J, Vitali F, Berghout J, Aberasturi D, Li J, et al. Novel disease syndromes unveiled by integrative multiscale network analysis of diseases sharing molecular effectors and comorbidities. BMC Med Genomics 2018;11:112. https://doi.org/10.1186/s12920-018-0428-9.

[48] de Jong S, Vidler LR, Mokrab Y, Collier DA, Breen G. Gene-set analysis based on the pharmacological profiles of drugs to identify repurposing opportunities in schizophrenia. J Psychopharmacol (Oxf) 2016;30:826–30. https://doi.org/10.1177/0269881116653109.

[49] Ferrero E, Agarwal P. Connecting genetics and gene expression data for target prioritisation and drug repositioning. BioData Min 2018;11:7. https://doi.org/10.1186/s13040-018-0171-y.

[50] So H-C, Chau CK-L, Chiu W-T, Ho K-S, Lo C-P, Yim SH-Y, et al. Analysis of genome-wide association data highlights candidates for drug repositioning in psychiatry. Nat Neurosci 2017;20:1342–9. https://doi.org/10.1038/nn.4618.

[51] Luo Y, Zhao X, Zhou J, Yang J, Zhang Y, Kuang W, et al. A network integration approach for drug-target interaction prediction and computational drug repositioning from heterogeneous information. Nat Commun 2017;8:1–13. https://doi.org/10.1038/s41467-017-00680-8.

[52] Lotfi Shahreza M, Ghadiri N, Mousavi SR, Varshosaz J, Green JR. A review of network-based approaches to drug repositioning. Brief Bioinform 2018;19:878–92. https://doi.org/10.1093/bib/bbx017.

[53] Alaimo S, Pulvirenti A. Network-Based Drug Repositioning: Approaches, Resources, and Research Directions. In: Vanhaelen Q, editor. Comput. Methods Drug Repurposing, New York, NY: Springer; 2019, p. 97–113. https://doi.org/10.1007/978-1-4939-8955-3_6.

[54] Wu Z, Wang Y, Chen L. Network-based drug repositioning. Mol Biosyst 2013;9:1268–81. https://doi.org/10.1039/C3MB25382A.

[55] Yu D, Kim M, Xiao G, Hwang TH. Review of Biological Network Data and Its Applications. Genomics Inform 2013;11:200–10. https://doi.org/10.5808/GI.2013.11.4.200.

[56] Stringer S, Wray NR, Kahn RS, Derks EM. Underestimated Effect Sizes in GWAS: Fundamental Limitations of Single SNP Analysis for Dichotomous Phenotypes. PLoS ONE 2011;6. https://doi.org/10.1371/journal.pone.0027964.

[57] So H-C, Chau CK-L, Lau A, Wong S-Y, Zhao K. Translating GWAS findings into therapies for depression and anxiety disorders: gene-set analyses reveal enrichment of psychiatric drug classes and implications for drug repositioning. Psychol Med 2018:1–17. https://doi.org/10.1017/S0033291718003641.

[58] Bakshi A, Zhu Z, Vinkhuyzen AAE, Hill WD, McRae AF, Visscher PM, et al. Fast set-based association analysis using summary data from GWAS identifies novel gene loci for human complex traits. Sci Rep 2016;6:32894. https://doi.org/10.1038/srep32894.

[59] Leeuw CA de, Mooij JM, Heskes T, Posthuma D. MAGMA: Generalized Gene-Set Analysis of GWAS Data. PLOS Comput Biol 2015;11:e1004219.

https://doi.org/10.1371/journal.pcbi.1004219.
[60] Jhamb D, Magid-Slav M, Hurle MR, Agarwal P. Pathway analysis of GWAS loci identifies novel drug targets and repurposing opportunities. Drug Discov Today 2019;24:1232–6. https://doi.org/10.1016/j.drudis.2019.03.024.
[61] Shen J, Song K, Slater AJ, Ferrero E, Nelson MR. STOPGAP: a database for systematic target opportunity assessment by genetic association predictions. Bioinformatics 2017;33:2784–6. https://doi.org/10.1093/bioinformatics/btx274.
[62] Gaspar HA, Breen G. Drug enrichment and discovery from schizophrenia genome-wide association results: an analysis and visualisation approach. Sci Rep 2017;7:1–9. https://doi.org/10.1038/s41598-017-12325-3.
[63] Wong BC, Chau CK, Ao F, Mo C, Wong S, Wong Y, et al. Differential associations of depression-related phenotypes with cardiometabolic risks: Polygenic analyses and exploring shared genetic variants and pathways. Depress Anxiety 2019;36:330–44. https://doi.org/10.1002/da.22861.
[64] So H-C, Wong Y-H. Implications of de novo mutations in guiding drug discovery: A study of four neuropsychiatric disorders. J Psychiatr Res 2019;110:83–92. https://doi.org/10.1016/j.jpsychires.2018.12.015.
[65] Breen G, Li Q, Roth BL, O'Donnell P, Didriksen M, Dolmetsch R, et al. Translating genome-wide association findings into new therapeutics for psychiatry. Nat Neurosci 2016;19:1392–6. https://doi.org/10.1038/nn.4411.
[66] Napolitano F, Zhao Y, Moreira VM, Tagliaferri R, Kere J, D'Amato M, et al. Drug repositioning: a machine-learning approach through data integration. J Cheminformatics 2013;5:30. https://doi.org/10.1186/1758-2946-5-30.
[67] Gottlieb A, Stein GY, Ruppin E, Sharan R. PREDICT: a method for inferring novel drug indications with application to personalized medicine. Mol Syst Biol 2011;7:496. https://doi.org/10.1038/msb.2011.26.
[68] Wang H, Gu Q, Wei J, Cao Z, Liu Q. Mining drug–disease relationships as a complement to medical genetics-based drug repositioning: Where a recommendation system meets genome-wide association studies. Clin Pharmacol Ther 2015;97:451–4. https://doi.org/10.1002/cpt.82.
[69] Lamb J, Crawford ED, Peck D, Modell JW, Blat IC, Wrobel MJ, et al. The Connectivity Map: Using Gene-Expression Signatures to Connect Small Molecules, Genes, and Disease. Science 2006;313:1929–35. https://doi.org/10.1126/science.1132939.
[70] Keenan AB, Jenkins SL, Jagodnik KM, Koplev S, He E, Torre D, et al. The Library of Integrated Network-Based Cellular Signatures NIH Program: System-Level Cataloging of Human Cells Response to Perturbations. Cell Syst 2018;6:13–24. https://doi.org/10.1016/j.cels.2017.11.001.
[71] Musa A, Ghoraie LS, Zhang S-D, Glazko G, Yli-Harja O, Dehmer M, et al. A review of connectivity map and computational approaches in pharmacogenomics. Brief Bioinform 2017;19:506–23. https://doi.org/10.1093/bib/bbw112.
[72] Barbeira AN, Dickinson SP, Bonazzola R, Zheng J, Wheeler HE, Torres JM, et al. Exploring the phenotypic consequences of tissue specific gene expression variation inferred from GWAS summary statistics. Nat Commun 2018;9:1–20. https://doi.org/10.1038/s41467-018-03621-1.
[73] Langhans SA. Three-Dimensional in Vitro Cell Culture Models in Drug Discovery and Drug Repositioning. Front Pharmacol 2018;9. https://doi.org/10.3389/fphar.2018.00006.

[74] Ioannidis NM, Wang W, Furlotte NA, Hinds DA, Bustamante CD, Jorgenson E, et al. Gene expression imputation identifies candidate genes and susceptibility loci associated with cutaneous squamous cell carcinoma. Nat Commun 2018;9:1–9. https://doi.org/10.1038/s41467-018-06149-6.
[75] Xue H, Li J, Xie H, Wang Y. Review of Drug Repositioning Approaches and Resources. Int J Biol Sci 2018;14:1232–44. https://doi.org/10.7150/ijbs.24612.
[76] Brohée S, Faust K, Lima-Mendez G, Vanderstocken G, Helden J van. Network Analysis Tools: from biological networks to clusters and pathways. Nat Protoc 2008;3:1616–29. https://doi.org/10.1038/nprot.2008.100.
[77] Sharma A, Ali HH. Analysis of clustering algorithms in biological networks. 2017 IEEE Int. Conf. Bioinforma. Biomed. BIBM, 2017, p. 2303–5. https://doi.org/10.1109/BIBM.2017.8218036.
[78] Picart-Armada S, Barrett SJ, Willé DR, Perera-Lluna A, Gutteridge A, Dessailly BH. Benchmarking network propagation methods for disease gene identification. PLOS Comput Biol 2019;15:e1007276. https://doi.org/10.1371/journal.pcbi.1007276.
[79] Luo H, Wang J, Li M, Luo J, Peng X, Wu F-X, et al. Drug repositioning based on comprehensive similarity measures and Bi-Random walk algorithm. Bioinformatics 2016;32:2664–71. https://doi.org/10.1093/bioinformatics/btw228.
[80] Ma X, Gao L. Biological network analysis: insights into structure and functions. Brief Funct Genomics 2012;11:434–42. https://doi.org/10.1093/bfgp/els045.
[81] Ideker T, Nussinov R. Network approaches and applications in biology. PLOS Comput Biol 2017;13:e1005771. https://doi.org/10.1371/journal.pcbi.1005771.
[82] Charitou T, Bryan K, Lynn DJ. Using biological networks to integrate, visualize and analyze genomics data. Genet Sel Evol 2016;48:27. https://doi.org/10.1186/s12711-016-0205-1.
[83] Zhu X, Gerstein M, Snyder M. Getting connected: analysis and principles of biological networks. Genes Dev 2007;21:1010–24. https://doi.org/10.1101/gad.1528707.
[84] Gaspar HA, Gerring Z, Hübel C, Major Depressive Disorder Working Group of the Psychiatric Genomics Consortium, Middeldorp CM, Derks EM, et al. Using genetic drug-target networks to develop new drug hypotheses for major depressive disorder. Transl Psychiatry 2019;9:117. https://doi.org/10.1038/s41398-019-0451-4.
[85] Ghosh Sujoy, Vivar Juan, Nelson Christopher P., Willenborg Christina, Segrè Ayellet V., Mäkinen Ville-Petteri, et al. Systems Genetics Analysis of Genome-Wide Association Study Reveals Novel Associations Between Key Biological Processes and Coronary Artery Disease. Arterioscler Thromb Vasc Biol 2015;35:1712–22. https://doi.org/10.1161/ATVBAHA.115.305513.
[86] Shu L, Chan KHK, Zhang G, Huan T, Kurt Z, Zhao Y, et al. Shared genetic regulatory networks for cardiovascular disease and type 2 diabetes in multiple populations of diverse ethnicities in the United States. PLOS Genet 2017;13:e1007040. https://doi.org/10.1371/journal.pgen.1007040.
[87] Greene CS, Krishnan A, Wong AK, Ricciotti E, Zelaya RA, Himmelstein DS, et al. Understanding multicellular function and disease with human tissue-specific networks. Nat Genet 2015;47:569–76. https://doi.org/10.1038/ng.3259.
[88] Dai Y-F, Zhao X-M. A Survey on the Computational Approaches to Identify Drug Targets in the Postgenomic Era. BioMed Res Int 2015. https://doi.org/10.1155/2015/239654.
[89] Radivojac P, Clark WT, Oron TR, Schnoes AM, Wittkop T, Sokolov A, et al. A large-

scale evaluation of computational protein function prediction. Nat Methods 2013;10:221–7. https://doi.org/10.1038/nmeth.2340.

[90] Wu Z, Wang Y, Chen L. Drug repositioning framework by incorporating functional information. IET Syst Biol Stevenage 2013;7:188–94.

[91] Burgess S, Small DS, Thompson SG. A review of instrumental variable estimators for Mendelian randomization. Stat Methods Med Res 2017;26:2333–55. https://doi.org/10.1177/0962280215597579.

[92] Mokry LE, Ahmad O, Forgetta V, Thanassoulis G, Richards JB. Mendelian randomisation applied to drug development in cardiovascular disease: a review. J Med Genet 2015;52:71–9. https://doi.org/10.1136/jmedgenet-2014-102438.

[93] Walker VM, Davey Smith G, Davies NM, Martin RM. Mendelian randomization: a novel approach for the prediction of adverse drug events and drug repurposing opportunities. Int J Epidemiol 2017;46:2078–89. https://doi.org/10.1093/ije/dyx207.

[94] Hon-Cheong S, Chau CK, Zhao K. Exploring repositioning opportunities and side-effects of statins: a Mendelian randomization study of HMG-CoA reductase inhibition with 55 complex traits. Genetics; 2017. https://doi.org/10.1101/170241.

[95] Zhao K, So H-C. Using Drug Expression Profiles and Machine Learning Approach for Drug Repurposing. In: Vanhaelen Q, editor. Comput. Methods Drug Repurposing, New York, NY: Springer; 2019, p. 219–37. https://doi.org/10.1007/978-1-4939-8955-3_13.

[96] Vamathevan J, Clark D, Czodrowski P, Dunham I, Ferran E, Lee G, et al. Applications of machine learning in drug discovery and development. Nat Rev Drug Discov 2019;18:463–77. https://doi.org/10.1038/s41573-019-0024-5.

[97] Chen H, Engkvist O, Wang Y, Olivecrona M, Blaschke T. The rise of deep learning in drug discovery. Drug Discov Today 2018;23:1241–50. https://doi.org/10.1016/j.drudis.2018.01.039.

[98] Yella J, Yaddanapudi S, Wang Y, Jegga A. Changing Trends in Computational Drug Repositioning. Pharmaceuticals 2018;11:57. https://doi.org/10.3390/ph11020057.

[99] Zhao K, So H-C. Drug Repositioning for Schizophrenia and Depression/Anxiety Disorders: A Machine Learning Approach Leveraging Expression Data. IEEE J Biomed Health Inform 2019;23:1304–15. https://doi.org/10.1109/JBHI.2018.2856535.

[100] Dugger SA, Platt A, Goldstein DB. Drug development in the era of precision medicine. Nat Rev Drug Discov 2018;17:183–96. https://doi.org/10.1038/nrd.2017.226.

[101] Yin L, Chau CKL, Sham P-C, So H-C. Uncovering complex disease subtypes by integrating clinical data and imputed transcriptome from genome-wide association studies: Applications in psychiatry and cardiovascular medicine. 2019 Am J Hum Genet, In Press. Available at BioRxiv. https://doi.org/10.1101/595488.

**1.**

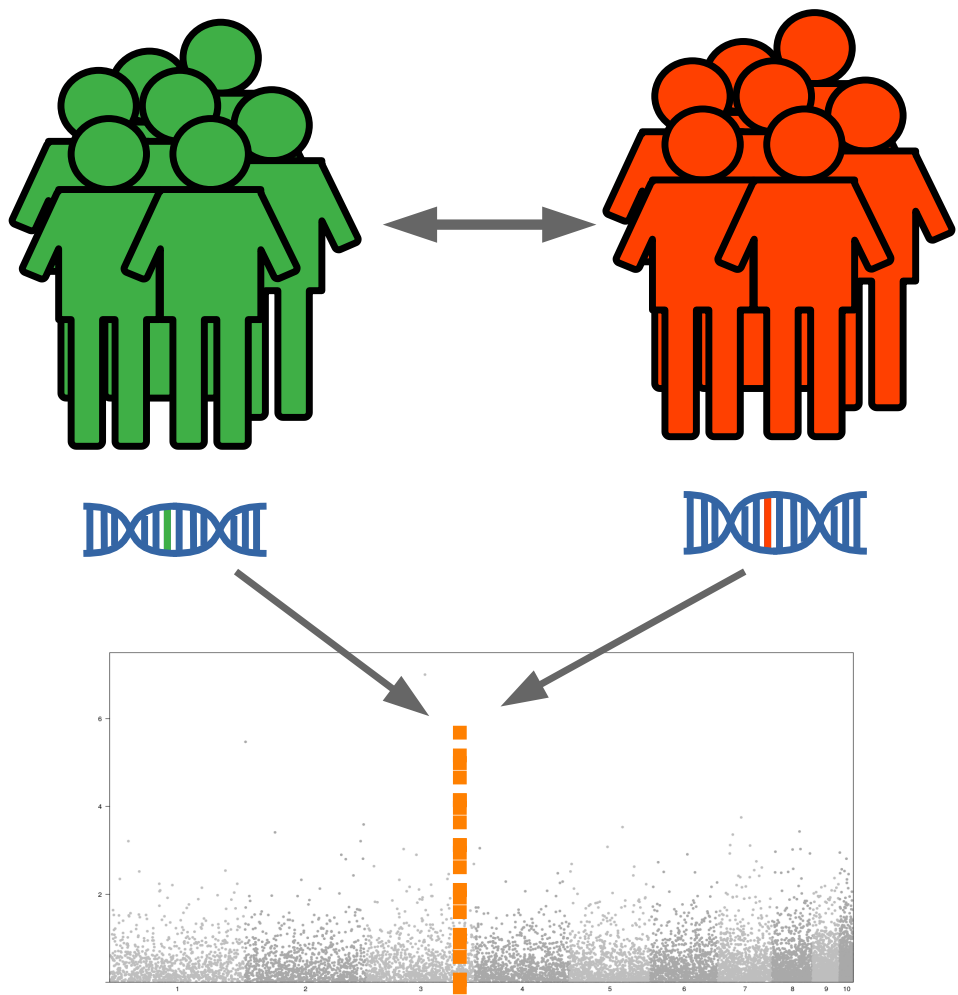

**2.**

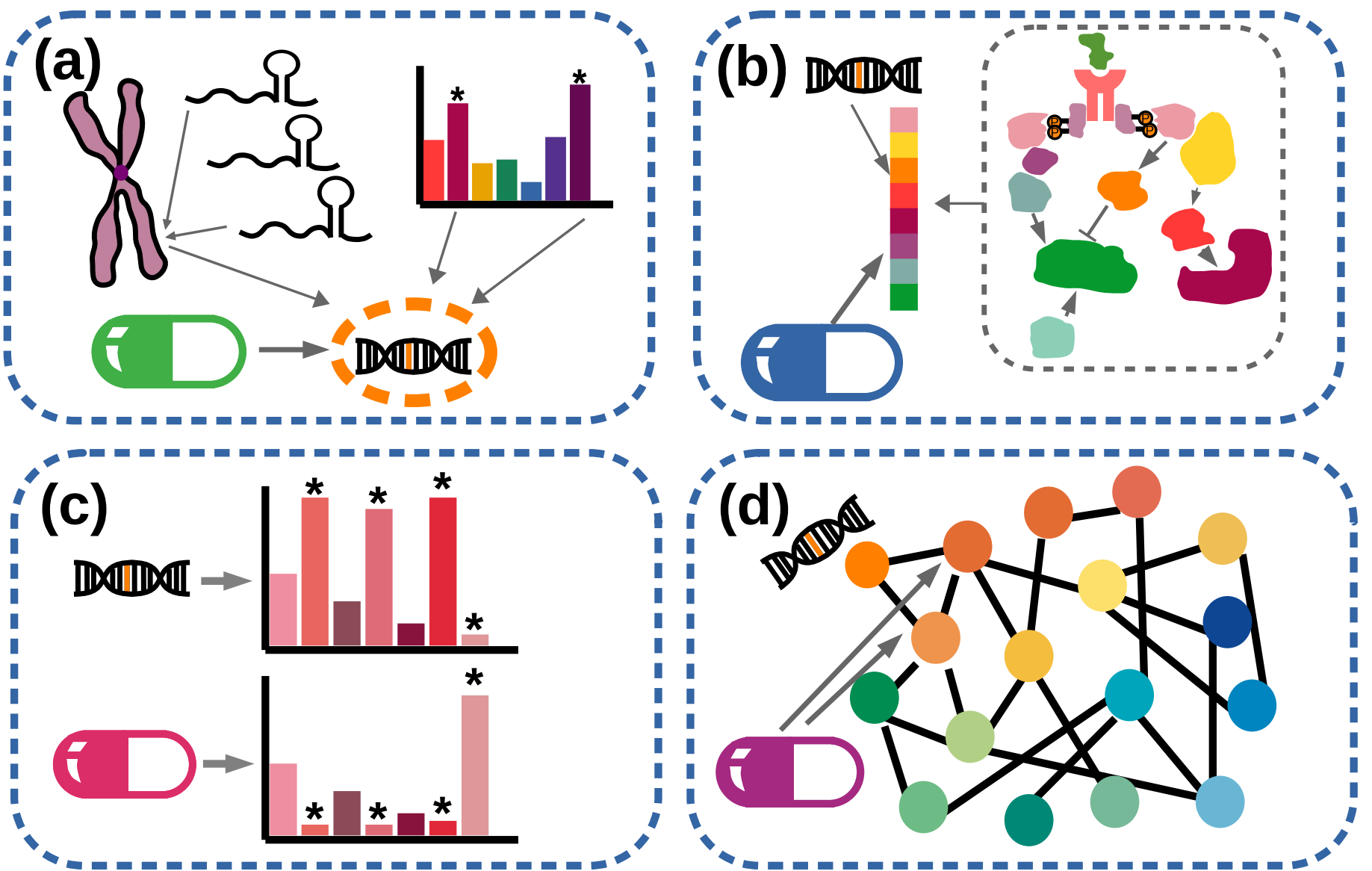

**3.**

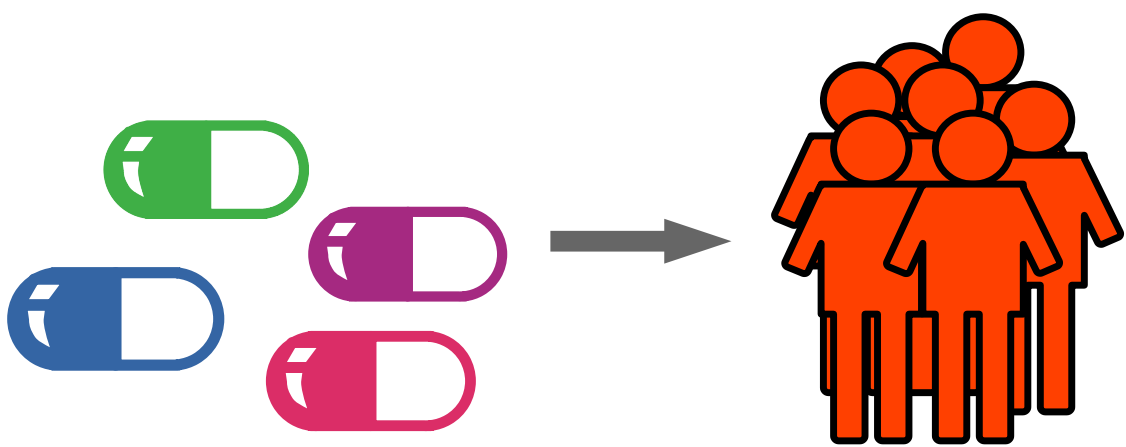